\documentstyle[aps,prl,multicol,epsf]{revtex}
\begin{document}

\newcommand{\parl}{\mbox{\tiny $| \! |$}}
\newcommand{\hide}[1]{}


\title{Dephasing at Low Temperatures}
\author{Doron Cohen and Yoseph Imry}

\date{June 1998, September 1998}

\address
{
Department of Physics,  
The Weizmann Institute of Science, Rehovot 76100, Israel. 
} 

\maketitle


\begin{abstract}
We discuss the significance and the calculation of 
dephasing at low temperatures. The particle is 
moving diffusively due to a static disorder configuration,  
while the interference between classical paths is suppressed 
due to the interaction with a dynamical environment.
At high temperatures we may use the `white noise 
approximation' (WNA), while at low temperatures we distinguish 
the contribution of `zero point fluctuations' (ZPF) from the  
`thermal noise contribution' (TNC). We study the limitations 
of the above semiclassical approach and suggest the 
required modifications. In particular we find that the ZPF 
contribution becomes irrelevant for thermal motion.   
\end{abstract}

\begin{multicols}{2}

The application of semiclassical considerations into the 
analysis of interference and dephasing is as old as the 
history of quantum mechanics. A particular interest is 
to apply these considerations to the theory of diffusing 
electrons in a metal \cite{AAK,imry}. 
Dephasing, or the loss of 'phase' information, 
is the consequence of the interaction with 
some other environmental degrees of freedom, or with some 
measurement device. The widely accepted dogma is that 
dephasing is associated with {\em leaving a trace} in the environment. 
Lately this dogma has been challenged experimentally \cite{webb} 
as well as theoretically \cite{zaikin}, leading to an intensive 
debate \cite{amb,alt} associated with the question whether 
ZPF of environmental modes may lead 
to decoherence at the limit of zero temperature.  
The most physically-appealing theoretical considerations are 
based on the application of Feynman-Vernon (FV) formalism, 
and simple semiclassical considerations. In particular, 
the zero temperature decoherence found within the exactly 
solvable Caldeira-Leggett (CL) model, is most puzzling. 
Thus, it is of fundamental importance to address the following: 
{\bf (1)} What are the limitations of the semiclassical strategy; 
{\bf (2)} What is the applicability of results that are based on 
the CL model; {\bf (3)} What are the limitations of the 'one particle' 
picture when applied to a many-body system.      
In this letter we are going to explore these questions
systematically, using a well controlled strategy.  
Under consideration is the motion of a particle under the 
combined influence of static disorder and a dynamical  
environment. Our reasoning consist of the following three stages:   
{\bf (a)} We take the static disorder into account 
and write the transport probability amplitude 
as a discrete sum over classical trajectories; 
{\bf (b)} We take the stochastic nature of the environment 
into account by considering the influence of 
an effective stochastic potential;     
{\bf (c)} We take the full dynamical nature of the environment 
into account by using the FV formalism.  
Having an environment whose temperature is~$T$, and a particle 
that is injected with an energy~$E$ (not necessarily thermal), 
one should make a distinction between the cases of high and low 
temperature~$T$, and analogous distinction between the cases 
of large and small energy~$E$. We shall argue that the validity of 
the semiclassical approach is not restricted to high bath-temperatures. 
However, at low  bath-temperatures, an essential modification 
is required in case of motion with small energy~$E$.

It is assumed that the motion of the particle 
under the influence of the static disorder 
is diffusive. The transport probability amplitude
can be written as a sum $\sum A_a \exp(iS[\bbox{x}_a]/\hbar)$
over classical trajectories that connect the observation point  
with the injection point. We shall denote by $t$ the total
time of the motion, and $S[\bbox{x}_a]$ are the corresponding 
action integrals. The statistical properties of 
the differences $\bbox{x}_b(t_2){-}\bbox{x}_a(t_1)$, 
where $a\&b$ is a pair of trajectories,   
will play a major role in later calculations. 
These statistical properties can be taken into 
account via a single function $P(k,\omega)$ that reflects 
the power-spectrum of the motion \cite{qbm}.  
For diffusive motion a practical approximation is   
\begin{eqnarray} \label{e1}
P(\bbox{k},\omega) \ = \ {2Dk^2} \ / \ {((Dk^2+1/t)^2+\omega^2)}
\end{eqnarray}
where $D$ is the diffusion coefficient. 
In the large $k$ regime where $k{>}(Dt)^{-1/2}$ this function 
is a properly normalized Lorentzian. Note that 
the collision frequency $\omega{\sim}v^2/D$, where $v$ is 
the velocity of the particle, should be used as a cutoff 
to the slow $1/\omega^2$ power-law decay. 
Beyond $k{\sim}v/D$ the above expression is not valid and 
the power spectrum of the motion is ballistic-like.

We now take the stochastic nature of the environment 
into account by introducing into the Hamiltonian 
a stochastic potential that satisfies
\begin{eqnarray} \label{e3} 
\langle{\cal U}(\bbox{x}'',t''){\cal U}(\bbox{x}',t')\rangle
=\phi(t''{-}t')\cdot w(\bbox{x}''{-}\bbox{x}') 
\end{eqnarray}
It is assumed that $w(\bbox{r})$ depends only on $|\bbox{r}|$.
The intensity of the noise is characterized 
by the parameter 
\begin{eqnarray}   \label{e4}
\nu \equiv \int_{-\infty}^{\infty}\phi(\tau)d\tau \cdot |w''(0)|
\end{eqnarray}
The power spectrum of the noise $\phi(\omega)$ is the 
Fourier transform of $\phi(\tau)$. We shall assume 
{\em ohmic environment}, meaning that at the classical limit 
$\phi(\omega)=\nu$ up to some cutoff frequency $1/\tau_c$ 
which is assumed to be larger than any other relevant 
frequency scale. Thus, in the classical limit we can use 
the WNA, namely $\phi(\tau)=\nu\delta(\tau)$. 
For the quantum mechanical see (\ref{e10}).  
Without loss of generality we shall assume the 
normalization $w''(0){=}-1$. The \mbox{$d$-dimensional}
Fourier transform of $w(\bbox{r})$ will be denoted by 
$\tilde{w}(\bbox{k})$. The mode-density 
(after angular integration) 
is $g(k) = (C_d/(2\pi)^d) k^{d{-}1} \tilde{w}(k)$, 
where $C_d$ is the total solid angle.
We shall assume that
\begin{eqnarray} \label{e5}  
g(k)=C \ell^{2+\sigma} k^{\sigma{-}1}
\ \ \ \ \mbox{for} \ \ k<1/\ell
\end{eqnarray}
where $\ell$ characterize the spatial scale of the 
correlations, and $C$ is a dimensionless constant. 
In case of short range Gaussian-type 
correlations $\sigma$ equals simply $d$. For the 
long range Coulomb interaction to be discussed later 
it equals $d{-}2$. In order to have a well 
defined model we must have $|w''(0)|<\infty$ therefore
only $-2<\sigma$ is meaningful. The regime $-2{<}\sigma{\le}0$
is well defined but it requires special treatment since 
$w(0)$ diverges.

The path-integral expression for the probability to 
propagate from the injection point to the observation point  
constitutes a double sum 
$\int\int{\cal D}\bbox{x}'{\cal D}\bbox{x}''$ over the 
path variables $\bbox{x}'(\tau)$ and $\bbox{x}''(\tau)$.
Averaging over realizations of ${\cal{U}}$  
one obtains the influence functional $\exp(-S_N/\hbar^2)$.
See \cite{dld} for details. The noise action functional is
\begin{eqnarray} 
S_N[\bbox{x}',\bbox{x}'']=\frac{1}{2} \int_0^t\int_0^t dt_1 dt_2 
\ \phi(t_2{-}t_1) \ \times \ \ \nonumber \\
\label{e6} 
\ \ \ \ [w(\bbox{x}_2''{-}\bbox{x}_1'')+
w(\bbox{x}_2'{-}\bbox{x}_1')-2w(\bbox{x}_2''{-}\bbox{x}_1')]
\end{eqnarray}
where $\bbox{x}_i$ is a shorthand notation for $\bbox{x}(t_i)$.
The corresponding semiclassical expression for that probability is
\begin{eqnarray} \label{e7}     
\sum_{ab} A_aA_b^{*} 
\exp\left(-\frac{S_N[\bbox{x}_a,\bbox{x}_b]}{\hbar^2}\right) \ 
\exp\left(i\frac{S[\bbox{x}_a]{-}S[\bbox{x}_b]}{\hbar}\right)
\end{eqnarray}
It is obvious that the interference contribution 
(the terms $a {\ne} b$) is suppressed due 
to the noise, while the classical (diagonal) contribution 
survives~{\dag}. 
This is the the {\em dephasing} effect. See general 
discussion in~\cite{imry}.
We are interested here in the computation of these 
dephasing factors, as well as in illuminating their 
physical significance. For the purpose of    
calculating the typical value of the dephasing factor,  
(\ref{e6}) can be replaced by 
\begin{eqnarray} \label{e12}  
\langle S_N \rangle \ \approx \  
t \int_0^{\infty} g(k)dk
\int_0^{\infty} \frac{d\omega}{\pi}  \phi(\omega)  P(k,\omega)
\end{eqnarray}
If the integral on the right hand side is independent 
of $t$, then the typical dephasing factor in (\ref{e7}) 
can be written as a simple exponential $\exp(-t/\tau_{\varphi})$. 

The full analysis should take into account the dynamical 
nature of the environment. Namely, one should consider an Hamiltonian 
${\cal H}={\cal H}_0(\bbox{x},\bbox{p})+
{\cal H}_{env}(\bbox{x},Q_{\alpha},P_{\alpha})$, 
where the latter term incorporates the interaction with environmental degrees 
of freedom. It is possible in principle (but generally not in practice) 
to use the FV formalism in order to derive an 
appropriate influence functional $\exp(iS_F/\hbar-S_N/\hbar^2)$. 
The Fluctuation-Dissipation theorem (FDT) implies that if $S_N$ is known, 
and the temperature of the bath is further specified, then also 
some of the dissipative properties of the environment are determined 
uniquely. Therefore it is plausible that the CL procedure 
of constructing an effective harmonic-bath, is useful in order to derive 
an actual expression for the friction functional $S_F$.  Indeed, this 
strategy has been adapted in \cite{dld} and lead to the introduction
of the DLD model~{\ddag}. 
The interaction with the bath-oscillators is 
\begin{eqnarray}   \label{e_Hint}
{\cal H}_I = \sum _{\alpha} c_{\alpha} Q_{\alpha} 
u(\bbox{x}{-}\bbox{x}_{\alpha}) \ \ \ .
\end{eqnarray}
Here $x_{\alpha}$ is the (fixed) location 
of the ${\alpha}$ oscillator and  
$Q_{\alpha}$ is its dynamical coordinate.  
The bath-oscillators are distributed uniformly 
all over space. The interaction of the particle with 
each of the oscillators is described by $u(r)$. 
The range of the interaction is $\ell$,  
and $c_{\alpha}$ are coupling constants. 
For an ohmic bath the following expression 
(generalized here for any dimension) has been derived:
\begin{eqnarray}   \label{e8}
S_F \ = \ \eta \int_0^t d\tau \ 
\nabla w(\bbox{r}) \cdot \dot{\bbox{R}}
\end{eqnarray}
where $\eta$ is a friction parameter, and the path 
variables are $\bbox{r}=\bbox{x}''{-}\bbox{x}'$ 
and $\bbox{R}=(\bbox{x}''{+}\bbox{x}')/2$.
From FDT it follows that if an ohmic environment is 
characterized by a temperature $T$ then the friction 
parameter should be $\eta = {\nu}/{(2k_BT)}$
We shall assume an environment that is characterized by 
a short spatial autocorrelation scale $\ell$, such that 
the classical trajectories are well separated with respect 
to this microscopic scale. Under such circumstances it has 
been observed in \cite{dld} that $S_F$ will have no 
effect on the interference contribution.  This statement 
does not hold in case of the CL model.
The CL version for $S_F$ is obtained by taking 
in (\ref{e8}) the limit $\ell\rightarrow\infty$, 
which is equivalent to the formal substitution
$w(\bbox{r})=-\bbox{r}^2/2$. Averaging the factor $\exp(iS_F/\hbar)$
over diffusive trajectories one obtains, as in \cite{amb}, 
a non-generic factor $\exp(-(t/\tau_{\varphi})^2)$ where 
$1/\tau_{\varphi}=\eta D /\hbar$. This particular result 
turns out to be identical, up to a logarithmic factor, 
with the genuine result (\ref{e13}), to be discussed later. 
However, it is not consistent to use the CL version 
for $S_F$ in the present circumstances, and therefore 
the approach of \cite{amb} does not apply.   
With the above observations, our semiclassical strategy implies 
that $S_F$ of a generic environment has no consequence on the 
analysis of dephasing, and Eq.(\ref{e7}) is still valid.  
Our main conclusion below will be that this (semiclassically-based) 
statement {\em fails} for a low energy particle.

We turn now to discuss some actual results for the dephasing rate. 
For ohmic bath the symmetrized power spectrum of the noise is
\begin{eqnarray}   \label{e10}
\phi(\omega) = \eta|\omega| 
\ \hbar\coth\left(\frac{\hbar|\omega|}{2k_BT}\right)
\ \ \ \ \mbox{for} \ \ |\omega|<1/\tau_c
\end{eqnarray}
For high temperatures $1/\tau_c < k_BT/\hbar$, 
one can use the WNA. 
Substituting $\phi(\tau){=}2\eta k_BT$ into (\ref{e6}), 
one obtains \cite{dld} the universal high temperature result 
\begin{eqnarray} \label{e11}  
\left( \frac{1}{\tau_{\varphi}} \right)_{\mbox{\tiny WNA}} 
\ = \ \frac{2\eta k_BT\ell^2}{\hbar^2} 
\ \ \ \ \ \ \mbox{for $0<\sigma$} \ .
\end{eqnarray}
This result does not apply for $-2{<}\sigma{\le}0$ 
(for electrons $d{\le}2$), because $w(0)$ diverges. 
Still, using the WNA and doing some simple manipulations,  
one obtains  $S_N{=}2\eta k_BT\ell^{2{+}\sigma}(Dt)^{-\sigma/2}t$, 
leading to a dephasing factor of the type 
$\exp(-(t/\tau_{\varphi})^{(2{-}\sigma)/2})$,
where $1/\tau_{\varphi} \propto T^{2/(2{-}\sigma)}$ 
in agreement with the well known results \cite{AAK,imry}.

\ \\
\epsfysize=1.6in
\epsffile{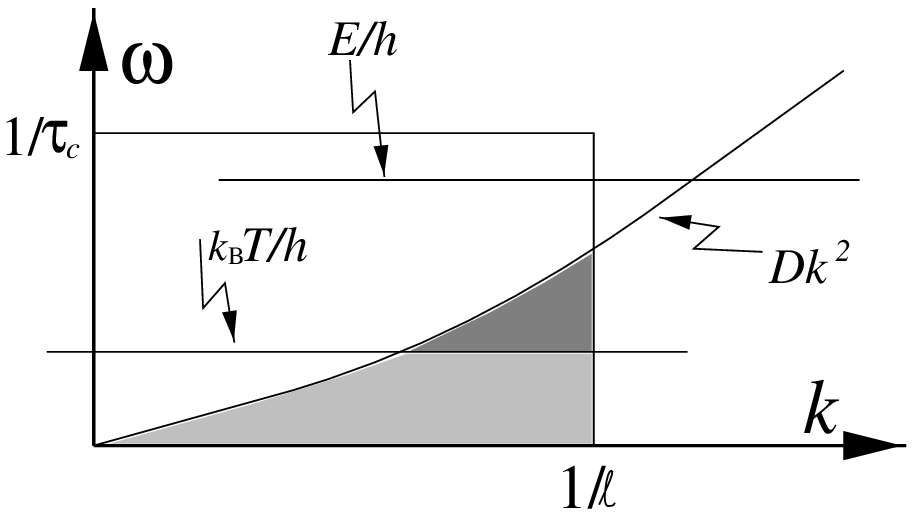}

\noindent
{\footnotesize {\bf FIG. 1.} 
The $(k,\omega)$ plane. The 
shaded regions indicate 
those environmental modes
that are effective in the 
dephasing process. The darker 
region indicates a possible 
excess contribution due to~ZPF.} 
\ \\

We are interested in going beyond the WNA. 
It is  useful to use (\ref{e12}) in order to 
perform the actual calculation.  
The domain of integration is illustrated in Fig.1. 
If the temperature is {\bf high} enough, such that 
$D/\ell^2<k_BT/\hbar$, then one can 
still use the WNA result. As the 
temperature becomes {\bf low}, such that 
$k_BT/\hbar<D/\ell^2$, the dephasing rate 
$1/\tau_{\varphi}$ becomes larger than the value 
which is predicted by the WNA.  
This is due to the ZPF in the frequency zone 
$k_BT/\hbar<\omega$ where  $\phi(\omega){=}\hbar\eta\omega$.
Later we shall see that $E/\hbar$ should be 
used as a cutoff for the $\omega$ integration. 
Therefore {\em the present analysis, and (\ref{e13}) below,  
does not apply to a low-energy particle}.   
Considering the ZPF contribution, the integral 
in (\ref{e12}) is dominated by $(k,\omega)$ modes that are 
concentrated along the curve $\omega=Dk^2$. 
Moreover, most of the contribution comes from 
modes with large wavenumber, namely $k {\sim} 1/\ell$. 
Thus the ZPF contribution to the dephasing rate  
is essentially as in \cite{zaikin} and applies 
at low temperatures to any $-2<\sigma$ 
\begin{eqnarray} \label{e13}  
\left( \frac{1}{\tau_{\varphi}} \right)_{\mbox{\tiny ZPF}} =  
\frac{C}{(2{+}\sigma)\pi}
\ln\left(1{+}\left(\frac{\ell v}{D}\right)^4\right)
\times\frac{1}{\hbar} \eta D \ . 
\end{eqnarray}
At low but finite temperatures, 
meaning $1/t < k_B T/\hbar \ll D/\ell^2$, 
one should consider the TNC that comes from 
the lower shaded region of Fig.1.  
For $-2{<}\sigma{\le}0$ the TNC dephasing 
factor is determined by $k{\sim}(Dt)^{-1/2}$ modes. 
For these modes $Dk^2{<}k_B T/\hbar$, 
and therefore the previously discussed WNA applies. 
On the other hand for $0{<}\sigma$ one obtains  
\begin{eqnarray} \label{e14}  
\left( \frac{1}{\tau_{\varphi}} \right)_{\mbox{\tiny TNC}} = 
{C'}\ell^{2{+}\sigma} 
\left(\frac{k_BT}{\hbar D}\right)^{\sigma/2}
\frac{2\eta k_BT }{\hbar^2} 
\end{eqnarray}
where $C'$ is a numerical factor.  
The above expression is valid for $0{<}\sigma{<}2$, 
where the dephasing process is dominated by modes 
with $k{\sim}(k_BT/\hbar D)^{1/2}$.   
For $2{<}\sigma$ the dephasing process is dominated 
by modes with $k{\sim}1/\ell$, and (\ref{e14}) 
should be modified by the replacement $\sigma \rightarrow 2$.      

The FV path-integral expression is {\em exact} in 
principle. On the other hand, our semiclassical 
expression (\ref{e7}) involves the stationary-phase 
{\em approximation}. Therefore it is important to 
understand, physically as well as mathematically, 
the validity limits of the semiclassical strategy. 
Each stationary-phase point of the exact FV path 
integral is a pair $a\&b$ of real classical trajectories 
that correspond  to the motion of a fictitious 
{\em classical test particle}. 
Let us assume for simplicity that $a\&b$ are loops 
related by time-reversal. It is also essential in the 
following argumentation that the interaction with the 
environment is {\em short range}, as in the 
$0{<}\sigma$ DLD model with $\ell$ which is a small scale. 
It follows from the definition of the influence 
functional that under such circumstances 
$P=1{-}\exp(-S_N[a])$ is the probability for a 
fictitious test particle to {\em leave a trace} 
along the way (i.e. to change the quantum mechanical 
state of at least one bath-oscillator along the loop).   
For simplicity one may consider the restricted 
problem of a particle that travels across a single 
bath-oscillator, meaning that (\ref{e_Hint}) includes 
only one term. It is well known $\cite{wisdom}$ that 
the semiclassical approximation is equivalent to 
a self-consistent mean-field scheme, where it is 
assumed that the wave-function for the 
particle-oscillator system can be written as 
a product of a scattered particle and driven oscillator.   
Such ansatz implies that for weak scattering we can treat 
the particle as moving with constant velocity $v$ and solve 
for the oscillator. It turns out that this reduction 
requires the assumption of small energy transfer.  
Therefore, one should anticipate problems once an oscillator 
with $\omega_{\alpha}$ larger than $E$ is involved. 
In the latter case, there is no justification to think 
of the particle as decoupled from the bath, moving 
with some constant velocity, capable of exciting 
the oscillator along the way. Therefore the 
corresponding probability $P$ loses its physical 
significance.  
Still, the interference contribution of the specified loop 
will be suppressed. It is true that a low-energy particle 
($\omega_{\alpha}<E$) is not capable of exciting the oscillator,  
but there is a finite probability to be scattered elastically, 
and consequently the original interference contribution 
is suppressed. However, this suppression does not have  
the meaning of genuine decoherence. Rather, it reflects  
a coherent ``re-normalization'' of the bare dynamics.    
One wonders whether $\exp(-S_N[a])$ gives an estimation  
for the  suppression of the original interference contribution 
due to these elastic scattering events. 
Unfortunately this is not the case. The elastic scattering 
off an oscillator involves a second-order process of virtual 
emission followed by absorption of a quanta $\hbar\omega_{\alpha}$. 
Therefore the elastic scattering probability is proportional 
to $c_{\alpha}^4$. At the same time $P$ is proportional 
to $c_{\alpha}^2$ in leading order.
Therefore, under such circumstances, $\exp(-S_N[a])$ does 
not reflect any physically meaningful quantity.  
It should be emphasized that the above argumentation 
{\em does not apply} once $\ell$ becomes large compared with 
the distance that is explored by the particle. 
In case of the CL model ($\ell\rightarrow\infty$) 
the decoherence process is no longer determined by 
the {\em scattering mechanism}. Rather, we have a crossover to 
a non-perturbative {\em spreading mechanism}. See detailed 
discussion of this point in \cite{qbm}.

The semiclassical strategy cannot be trusted if some of the 
effective bath-oscillators are such that $E/\hbar < \omega_{\alpha}$. 
For diffusive motion such a situation will occur 
if $E/\hbar < D/\ell^2$. 
Note that the notion of large/small energy is in complete 
analogy with the notion of high/low temperature. 
For thermal motion $E \sim k_BT$ and the two notions coincides.   
In order to see how expression (\ref{e12}) should be modified 
for small-energy motion, let us obtain it using an elementary 
perturbative calculation. This is straightforward for short range 
interaction ($0{<}\sigma$) since it is plausible that 
$1/\tau_{\varphi}$ has then the significance of inelastic 
scattering rate. The leading order inelastic scattering probability 
is expressed in terms of the diagonal matrix elements of the 
scattering matrix, namely 
\mbox{$P=\langle 1-|\langle\bbox{n}\psi|
{\cal S}|\bbox{n}\psi \rangle|^2\rangle_{\beta}$}.  
The subscript ${\beta}$ implies thermal average 
over the quantum-mechanical states $\bbox{n}$ of the 
bath-oscillators. The incoming particle is described 
by the wavefunction $\psi$. Note that the precise 
definition of `dephasing-rate' becomes vague once the 
semiclassical approach is abandoned. It seems 
plausible that $P$ should be averaged over eigenstates
of the disordered potential.      
Using second order perturbation theory  
\begin{eqnarray}  \nonumber 
\langle \bbox{n} \psi|S| \bbox{n} \psi\rangle =
1{-}\frac{1}{\hbar^2}\int_0^t\!\!\int_0^{t_2}\!\! dt_2dt_1 
\langle\bbox{n}\psi | 
{\cal H}_I(t_2) {\cal H}_I(t_1) 
| \bbox{n}\psi \rangle
\end{eqnarray}
Standard manipulations which are based on the  
definition of the ohmic DLD model lead to the 
result $P=1/\tau_{\varphi}$ where  
\begin{eqnarray} \label{e16}  
\frac{1}{\tau_{\varphi}} \ = \ \frac{1}{\hbar^2} 
\int \frac{\bbox{dk}}{(2\pi)^d} 
\int \frac{d\omega}{2\pi} 
\ \tilde{w}(k)\Phi(\omega) 
\ P_{qm}(-\bbox{k},-\omega)
\end{eqnarray}
Here $\Phi(\omega)=\hbar\eta|\omega|n(\omega)$ is the 
non-symmetrized version of $\phi(\omega)$, where 
$n(\omega)=1/(\exp(\hbar\omega/k_BT)-1)$ for $0{<}\omega$,  
and $n(\omega)=(1{+}n(|\omega|))$ for $\omega{<}0$.   
At zero temperature $\Phi(0{<}\omega){=}0$.
The quantum-mechanical power spectrum of the motion 
is defined as the (complex) Fourier transform of  
$P_{qm}(\bbox{k},\tau)$ which is the correlator 
of the operator $\exp(ik\bbox{x})$.   
Using the semiclassical estimate 
$P_{qm}(\bbox{k},\omega) \approx P(\bbox{k},\omega)$ 
one obtains again the integral expression in (\ref{e12}) 
for the dephasing rate. The validity of the semiclassical estimate
for $P_{qm}(\bbox{k},\omega)$ is restricted to the frequency range 
$|\omega|<E/\hbar$. Quantum-mechanics cannot support 
larger frequencies!  For thermal motion the 
result of the quantal FDT is related to 
the result of the classical FDT as follows:  
\begin{eqnarray} \label{e17}   
P_{qm}(\bbox{k},\omega) = 
\frac{\hbar|\omega|}{k_BT}n(\omega) \ P(\bbox{k},\omega)
\end{eqnarray}
Thus, it is suggested that for thermal motion $k_BT/\hbar$ 
should serve as an effective cutoff in the $\omega$ integration 
of Eq.(\ref{e12}).  Consequently the ZPF contribution 
for the dephasing rate {\em should be omitted}.  For non-thermal 
motion $E/\hbar$ is the proper cutoff and some (or all) of 
the ZPF contribution should be included.
 
Finally, one wonders whether additional modifications 
are required once we turn to treat (semiclassically) the 
problem of dephasing of electrons in a metal, 
taking into account the presence of Fermi sea.   
Here we consider ballistic-like motion as a test case. The 
average scattering rate can be calculated by using a kinetic picture. 
The transition rate for unit volume is 
$W(p'|p)=\tilde{w}(k)\Phi(\omega)/\hbar^2$, where $k{=}(p'{-}p)/\hbar$ 
and $\omega{=}(E(p'){-}E(p))/\hbar$. The average scattering rate is 
proportional to 
$\int\int\bbox{dp'}\bbox{dp}(1{-}f(E'))f(E)W(p'|p)$. Standard 
manipulation leads to Eq.(\ref{e16}) with (\ref{e17}). 
Note that for the ballistic-like motion under consideration 
one should substitute $P(\bbox{k},\omega)=1/(v|k|)$ for $|\omega|{<}v|k|$ 
and zero otherwise. 
This calculation implies that the dephasing rate is similar to the 
inelastic scattering rate also for the many-body case  
(provided $0{<}\sigma$).  

This research was supported by the Israel Science Foundation 
and by the German-Israeli Foundation~(GIF), Jerusalem. The 
authors are grateful to the Newton Institute Cambridge for 
hospitality when this work started. We thank Y.~Gefen,
D.E.~Khmelnitskii, Y.~Levinson, U.~Smilansky and A.~Stern 
for useful discussions.


\end{multicols}

\begin{thebibliography}{99}
\bibitem[\ddag]{ft2}  
DLD are the initials of "Diffusion Localization and 
Dissipation". Obviously, these three effects comes out naturally 
from the solution of the DLD model. Note however that localization 
effect is out of the scope of the preset application. See the 
previous footnote in this respect.
\end{thebibliography}
\end{document}